# Most stable superheavy nuclei in the island of stability

H.C.Manjunatha[1], L.Seenappa[1], K.N.Sridhar[2]
[1]Department of Physics, Government College for Women, Kolar-563101 Karnataka, INDIA
[2]Department of Physics, Government First Grade College, Kolar-563101 Karnataka, INDIA
**manjunathhc@rediffmail.com**

**Abstract**

*We have investigated most stable superheavy nuclei by studying the decay properties such as alpha decay, cluster decay and spontaneous fission. We have investigated nine stable nuclei in the island of stability which can be detected through fission are $^{318}$123(10.5ms), $^{319}$123(4.68μs), $^{317}$124($1.74\times10^{4}$ y), $^{318}$124($2.70\times10^{1}$ y), $^{319}$124($2.83\times10^{-2}$ y), $^{320}$124($1.91\times10^{-5}$ y), $^{319}$125($2.46\times10^{9}$ y), $^{320}$125($3.81\times10^{6}$ y) and $^{321}$125($3.99\times10^{3}$ y). Present work also investigates three stable superheavy nuclei which can be detected through alpha decay which are $^{318}$125($1.03\times10^{12}$ y), $^{319}$126($5.77\times10^{11}$ y) and $^{320}$126($3.99\times10^{10}$ y). These nuclei will become most stable nuclei if they synthesized in the laboratory. The identified twelve stable nuclei is the evidence for the hypothesis of island of stability.*

## I.Introduction

The unanswered questions in the field of Nuclear Physics are; what is the heaviest superheavy nuclei that can exist? and do very long-lived superheavy nuclei exist in the nature?. The past ten years have been marked by remarkable progress in the science of superheavy elements and nuclei. The island of stability was predicted theoretically more than 40 years ago. The lifetimes of most known superheavy nuclei are governed by the competition between α -decay and spontaneous fission. The existence of island of stability has been confirmed experimentally [1-3] during previous decade. Superheavy nuclei with Z=112–118 have successfully synthesized using fusion reactions with $^{48}$Ca beam and various actinide targets at FLNR (Dubna), GSI (Darmstadt), and LBNL (Berkeley) [4–12]. There were attempts made by the previous workers [13-15] to produce the nuclei with Z=119 and 120. Various phenomenological and microscopic models such as fission model [16], cluster model [17], generalized liquid drop model (GLDM) [18], unified model for alpha-decay and alpha capture (UMADAC) [19] are available in the literature to study the different decay modes of superheavy nuclei. simple empirical formulae [20-44] are also available to determining decay half-lives. Aim of the present work is to identify the stable superheavy nuclei, for this we have evaluated alpha decay halflives for the superheavy nuclei of atomic number range 104<Z<126.

## II. Theory

To identify the stable superheavy nuclei, we have investigated the alpha decay process using the following theoretical framework. The potential $V(R)$ is considered as the sum of the Coulomb, the nuclear and the centrifugal potentials

$$V(R) = V_C(R) + V_N(R) + V_{cf}(R) \qquad (1)$$

Coulomb potential $V_c(R)$ is taken as

$$V_C(R) = Z_1 Z_2 e^2 \begin{cases} \frac{1}{R} & (R > R_C) \\ \frac{1}{2R_c}\left[3 - \left(\frac{R}{R_c}\right)^2\right] & (R < R_C) \end{cases} \qquad (2)$$

where $R_C = 1.24 \times (R_1 + R_2)$, $R_1$and $R_2$ are respectively the radii of the emitted alpha and daughter nuclei. Here $Z_1$ and $Z_2$ are the atomic numbers of the daughter and emitted cluster the nuclear potential $V_N(R)$ is calculated from the proximity potential and it is given as

$$V_N(z) = 4\pi\gamma\left[\frac{C_1 C_2}{C_1 + C_2}\right]\Phi(\xi) \qquad (3)$$

We have used the Myers and Swiatecki [23] modified the proximity potential. This proximity potential is based on the droplet model concept. Using the droplet model, matter radius $C_i$ was calculated as

$$C_i = R_i + \frac{N_i}{A_i} t_i \qquad (4)$$

With $R_i(\alpha_i) = R_{0i}\left[1 + \sum_\lambda \beta_{\lambda i} Y_\lambda^{(0)}(\alpha_i)\right]$ (5)

Here $\alpha_i$ is the angle between the radius vector and symmetry axis of the $i^{th}$ nuclei and it is to be noted that the quadrupole interaction term proportional to $\beta_{21}\beta_{22}$ is neglected because of its short range character. For this potential, $R_{0i}$ denotes the half-density radii of the charge distribution and $t_i$ is the neutron skin of the nucleus. The nuclear charge, is given by the relation

$$R_{00i} = 1.24 A_i^{1/3}\left(1 + \frac{1.646}{A_i} - 0.191\frac{A_i - 2Z_i}{A_i}\right) fm \qquad (i=1,2) \qquad (6)$$

The half-density radius $c_i$ was obtained from the relation

$$R_{0i} = R_{00i}\left(1 - \frac{7}{2}\frac{b^2}{R_{00i}^2} - \frac{49}{8}\frac{b^4}{R_{00i}^4} + .....\right) \qquad (i=1,2) \qquad (7)$$

Using the droplet model, neutron skin $t_i$ reads as

$$t_i = \frac{3}{2} r_0 \left(\frac{J I_i - \frac{1}{12} c_1 Z_i A_i^{-1/3}}{Q + \frac{9}{4} J A_i^{-1/3}}\right) \qquad (i=1,2) \qquad (8)$$

here $r_0$ is 1.14 fm, the value of the nuclear symmetric energy coefficient $J$ = 32.65 MeV, $I_i = (N_i - A_i)/Z_i$ and $c_1 = 3e^2/5r_0$ = 0.757895 MeV. The neutron skin stiffness coefficient Q was taken to be 35.4MeV. The nuclear surface energy coefficient $\gamma$ in terms of neutron skin was given as

$$\gamma = \frac{1}{4\pi r_0^2}\left[18.63(MeV) - Q\frac{t_1^2 + t_2^2}{2 r_0^2}\right] \qquad (9)$$

where $t_1$ and $t_2$ were calculated using above equation. The universal function for this is given by

$$\Phi(\xi) = -0.1353 + \sum_{n=0}^{5} \frac{c_n}{(n+1)}(2.5 - \xi)^{n+1}, \text{ for } \mathbf{0 \le \xi \le 2.5} \qquad (10)$$

$$\Phi(\xi) = -exp\left(\frac{2.75-\xi}{0.7176}\right), \qquad \text{for } \xi \geq 2.5 \tag{11}$$

where $\xi = R - C_1 - C_2$. The values of different constants $c_n$ were $c_0$ =−0.1886, $c_1$ =−0.2628, $c_2$ =−0.15216, $c_3$ =−0.04562, $c_4$ =0.069136, and $c_5$ = −0.011454.
The Langer modified centrifugal barrier is adopted[44] in the present calculation

$$V_{cf} = \frac{h\left(l+\frac{1}{2}\right)^2}{4\pi \times \mu R^2} \tag{12}$$

According to WKB approximation (Wentzel-Kramers-Brillouin) the penetration probability P through the potential barrier studied by the following equation

$$P = \exp\left\{-\frac{2}{\eta}\int_{R_a}^{R_b}\sqrt{2\mu(V_T(r)-Q)}dr\right\} \tag{13}$$

where μ is the reduced mass alpha decay system, $R_a$ and $R_b$ are the inner and outer turning points and these turning points are calculated by

$$V_T(R_a) = Q = V_T(R_b) \tag{14}$$

The decay half-life of parent nuclei with the emission of alpha particle is studied by

$$T_{1/2} = \frac{\ln 2}{\lambda} = \frac{\ln 2}{\nu P} \tag{15}$$

Where λ is the decay constant and ν is the assault frequency and is expressed as

$$\nu = \frac{\omega}{2\pi} = \frac{2E_\nu}{h} \tag{16}$$

where $E_\nu$ is the empirical vibrational energy [39].

## III. Results and discussion

We have studied the alpha decay halflives of around 600 nuclei. The energy released ($Q_\alpha$) during the alpha decay between the ground-state of the parent nuclei and the ground-state of the daughter nuclei is calculated from the mass excess data of the parent and daughter nuclei [45-47]. The variation of mass excess of parent ($M_p$) and daughter ( $M_d$) during alpha decay with mass number (A) for superheavy elements of 104≤Z≤126 is as shown in figure 1. The energy released during the decay process is depend on the difference between the mass excess of parent and daughter. In the figure 1, we have highlighted the minimum energy difference ($M_p$-$M_d$) between parent and daughter. If the value of ($M_p$-$M_d$) is low then the corresponding half-lives are high. Figure 2 shows the variation of energy released Qα (MeV) during alpha decay with mass number (A) for superheavy elements of 104≤Z≤126. As the value of Qα is less then corresponding nuclei is having longer half-lives. In figure 2, we have highlighted the isotopes for each superheavy elements having minimum Q-value. It is identified that $Q_\alpha$ value is minimum for the nuclei $^{319}$123 (4.32MeV), $^{320}$124(4.13MeV), $^{321}$125 (4.27MeV) and $^{319}$126 (6.44MeV) which are located in the island of stability.

To investigate the dominant decay mode, we have also compared the alpha decay halflives with that of spontaneous fission. This comparison is as shown in figure 3. The superheavy nuclei synthesized till now (Z=104 to 118) are having half-lives from few seconds to μ seconds. It is surprisingly observed that the nuclei which lies in the island of stability such as $^{317-320}$124, $^{318-321}$125 and $^{319-320}$126 are having half-lives in years and appears to be stable among the superheavy region. To validate the present work, we have compared the alpha decay half-lives with that of the values produced by the different semi-empirical formulae in the literature [20-43]. Alpha decay half-lives produced by the present work agrees with that of semi-empirical formulae. This comparison is shown in table 1. To check the correctness of the method, we have also compared the logarithmic alpha decay half-lives produced by the present work with that of the experiments [5,13]. This comparison is as shown in table 2. From this comparison, it is found that resent work agree with the experiments. Among the identified stable nuclei lies in the island of stability, the nuclei $^{318}$125 $^{319-320}$126 sustain against fission. Whereas remaining stable nuclei in the island of stability $^{317-320}$124 and $^{319-321}$125 undergoes fission with half-lives in years.

After studying the alpha decay and fission properties, we have also studied the heavy particle radioactivity/cluster radioactivity ($^{9}$Be, $^{10}$B,$^{11}$B ,$^{12}$C, $^{14}$N, $^{16}$O, $^{19}$F, $^{20}$Ne, $^{21}$Ne, $^{22}$Ne, $^{23}$Na, $^{24}$Mg, $^{25}$Mg, $^{26}$Mg, $^{27}$Al, $^{28}$Si, $^{29}$Si, $^{30}$Si $^{31}$P $^{32}$S, $^{33}$S, $^{34}$S, $^{35}$Cl, $^{36}$Ar, $^{38}$Ar, $^{40}$Ar, $^{39}$K, $^{41}$K, $^{40}$Ca, $^{42}$Ca, $^{43}$Ca, $^{44}$Ca,$^{46}$Ca) in the identified stable nuclei $^{317-320}$124, $^{318-321}$125 and $^{319-320}$126 using the model explained in the previous work [48]. It is also found that the nuclei $^{317-320}$124, $^{318-321}$125 and $^{319-320}$126 are stable against cluster radioactivity and having halflives $10^{25}$ -$10^{70}$y. The stability of identified nuclei $^{317-320}$124, $^{318-321}$125 and $^{319-320}$126 are also checked against the proton, neutron, and beta emission by studying corresponding separation energies. It is also found that these nuclei are stable against neutron, proton, and beta decay. Table 3 presents the identified stable nuclei in the island of stability and corresponding decay modes. There are nine stable nuclei in the island of stability which can be detected through fission are $^{318}$123(10.5ms), $^{319}$123(4.68μs), $^{317}$124(1.74×$10^{4}$ y), $^{318}$124(2.70×$10^{1}$ y), $^{319}$124(2.83×$10^{-2}$ y), $^{320}$124(1.91×$10^{-5}$ y), $^{319}$125(2.46×$10^{9}$ y), $^{320}$125(3.81×$10^{6}$ y) and $^{321}$125(3.99×$10^{3}$ y). There are three stable superheavy nuclei which can be detected through alpha decay are $^{318}$125(1.03×$10^{12}$ y), $^{319}$126(5.77×$10^{11}$ y) and $^{320}$126(3.99×$10^{10}$ y).

Isotopes on the island are believed to have magic numbers of protons and neutrons that can have relatively long half-life to maintain stability. on par with this argument, the 12 nuclei identified in this work is having relatively long half-life and becomes stable nuclei among the superheavy elements. The identified twelve stable nuclei is the evidence for the hypothesis of island of stability.

## Conclusion

We have investigated nine stable nuclei in the island of stability which can be detected through fission are $^{318}$123(10.5ms), $^{319}$123(4.68μs), $^{317}$124(1.74×$10^{4}$ y), $^{318}$124(2.70×$10^{1}$ y), $^{319}$124(2.83×$10^{-2}$ y), $^{320}$124(1.91×$10^{-5}$ y), $^{319}$125(2.46×$10^{9}$ y), $^{320}$125(3.81×$10^{6}$ y) and $^{321}$125(3.99×$10^{3}$ y). Present work also investigates three stable superheavy nuclei which can be detected through alpha decay which are $^{318}$125(1.03×$10^{12}$ y), $^{319}$126(5.77×$10^{11}$ y) and $^{320}$126(3.99×$10^{10}$ y).

## Reference

[1] Yu. Ts. Oganessian, J. Phys. G:Nucl. Part.Phys. 34, R165, (2007)
[2] J. H. Hamilton, S.Hofmann, Y. T. Oganessian, Annu, Rev.Nucl. Part. Sci. 63, 383, (2013).
[3] G.G.Adamian, N.V.Antonenko, H.Lenske, Nuclear Physics A 970 (2018) 22–28

[4] Yu.Ts. Oganessian, et al., Phys. Rev. Lett. 104 (2010) 142502
[5] Yu.Ts. Oganessian, et al., Phys. Rev. C 87 (2013) 014302.
[6] V.K. Utyonkov, et al., Phys. Rev. C 92 (2015) 034609.
[7] Yu.Ts. Oganessian, V.K. Utyonkov, Nucl. Phys. A 944 (2015) 62.
[8] L. Stavsetra, K.E. Gregorich, J. Dvorak, P.A. Ellison, I. Dragojevic, M.A. Garcia, H. Nitsche, Phys. Rev. Lett. 103 (2009) 132502.
[9] Ch. Düllmann, et al., Phys. Rev. Lett. 104 (2010) 252701.
[10] J.M. Gates, et al., Phys. Rev. C 83 (2011) 054618.
[11] S. Hofmann, et al., Eur. Phys. J. A 48 (2012) 62.
[12] J.M. Khuyagbaatar, et al., Phys. Rev. Lett. 112 (2014) 172501
[13] Yu.Ts. Oganessian, et al., Phys. Rev. C 79 (2009) 024603.
[14] C.E. Düllmann, TASCA Collaboration, in: Fission and Properties of Neutron-Rich Nuclei, vol.44, World Scientific, Singapore, 2013, p.271.
[15] S. Hofmann, et al., Eur. Phys. J. A 52 (2016) 180, 52 (2016) 116.
[16] D.N. Poenaru, M. Ivascu, A. Sandulescu, W. Greiner, Phys. Rev. C 32 (1985)572.
[17] B. Buck, A.C. Merchant, S.M. Perez, Phys. Rev. C 45 (1992) 2247.
[18] H.F. Zhang, G. Royer, Phys. Rev. C 76 (2007) 047304.
[19] V. Yu. Denisov and H. Ikezoe, Phys. Rev. C 72 (2005) 064613
[20] D. N. Poenaru, R. A. Gherghescu, and W. Greiner, Phys. Rev. C **83**, 014601 (2011).
[21] D.Ni, Z.Ren, T.Dong, C.Xu, Phys. Rev. C 78, 044310 (2008)
[22] A. Sobiczewski, Z. Patyk, S. Cwiok, Phys.Lett. 224, 1 (1989)
[23] W.D. Myers and W.J. Swiatecki, Phys. Rev. C **62**, 044610 (2000).
[24] C. Qi, F. R. Xu, R. J. Liotta, and R. Wyss, Phys. Rev. Lett. **103**, 072501 (2009).
[25] B. Sahu, R. Paira, and B. Rath, Nucl. Phys. A 908, 40 (2013).
[26] Denisov, V. Y., & Khudenko, A. A. (2009). Phys. Rev. C, 79(5), 054614
[27] D.T.Akrawy, H.Hassanabadi, S.S.Hosseini, K.P.Santhosh, Nucl. Phys. A 971 (2018)130-137
[28] M.Horoi, B.A.Brown, A.Sandulescu, J. Phys. G: Nucl. Part. Phys. 30, 945 (2004)
[29] V.Yu.Denisov, A.A.Khudenko, Phys. Rev. C 79, 054614 (2009) 82, 059901 (E) (2010)
[30] A.Sobiczeweski, Z.Patyk, S.Cwiok, Phys.Lett B, 224(1989)1
[31] A.Parkhomenko, A.Sobiczeweski, Acta Physica Pol.B 36(10)(2005)3095-3108
[32] G.Royer, J. Phys. G: Nucl. Part. Phys. 26, 1149 (2000)
[33] A.Parkhomenko, A.Sobiczewski, Acta Physica pol. B 36(10)(2005) 3095-3108
[34] D.T.Akrawy, H.Hassanabadi, S.S.Hosseini, K.P.Santhosh, Nucl. Phys. A 971 (2018)130-137
[35] G.Royer, Nucl. Phys. A, 848, 279 (2010).
[36] B. Alex Brown, Phys. Rev. C 46, 811 (1992).
[37] B.A. Brown, Phys. Rev. C 46 (1992) 811.
[38] A.I. Budaca, R.Budaca, I.Silisteanu, Nuc. Phys. A 951 (2016) 60-74.
[39] D. N. Poenaru, W. Greiner, M. Ivaşcu, D. Mazilu, and I. H. Plonski, Z. Phys. A: At. Nucl. 325, 435 (1986).
[40] A.Sobiczewski, A.Parkhomenko, Prog. Part. Nucl. Phys. 58, 292 (2007)
[41] D.T. Akrawy and D. N. Poenaru, Jour. Phys. G: Nuc. and Part. Phys. 44, 10 (2017).
[42] D. Akrawy and D. N. Poenaru, J. Phys. G: Nucl. Part. Phys. 44, 105105 (2017).
[43] J. M. Dong, et al., Nucl. Phys. A 832, 198 (2010)

[44] M. Ismail, A. Y. Ellithi, M. M. Botros and A. Adel, Phys. Rev. C 81 (2010) 024602.
[45] H.C. Manjunatha (2016) Int. J Mod. Phy. E 25(11):1650100
[46] H.C. Manjunatha (2016) Int. J Mod. Phy. E 25(9):1650074.
[47] H.C. Manjunatha (2016) *Nucl. Phy. A (2016)* 945:42–57
[48] H.C. Manjunatha, N.Sowmya (2018) *Nucl. Phy. A (2018)* 969: *68–82*

Table 1: comparison of the alpha decay halflives with that of the values produced by the different semi empirical formulae available in the literature

| Method | $^{318}$**123** | $^{319}$**123** | $^{317}$**124** | $^{318}$**124** | $^{319}$**124** | $^{320}$**124** | $^{318}$**125** | $^{319}$**125** | $^{320}$**125** | $^{321}$**125** | $^{319}$**126** | $^{320}$**126** |
|---|---|---|---|---|---|---|---|---|---|---|---|---|
| Present work | 32.87 | 35.95 | 15.96 | 29.00 | 34.51 | 38.18 | 17.18 | 29.95 | 34.87 | 37.14 | 19.26 | 18.10 |
| UNIV [20] | 32.49 | 34.97 | 15.61 | 28.61 | 34.16 | 37.83 | 16.86 | 29.44 | 34.54 | 36.78 | 18.89 | 17.71 |
| NRDX[21] | 33.69 | 36.06 | 16.54 | 28.85 | 34.29 | 37.62 | 18.76 | 30.81 | 35.67 | 37.80 | 19.73 | 18.46 |
| TNF [22] | 24.24 | 26.64 | 7.77 | 20.44 | 25.81 | 29.35 | 8.96 | 21.21 | 26.15 | 28.32 | 10.91 | 9.77 |
| UDL1[23] | 32.22 | 34.61 | 15.82 | 28.48 | 33.84 | 37.37 | 17.05 | 29.30 | 34.22 | 36.37 | 19.05 | 17.89 |
| UDL2[24] | 32.17 | 34.49 | 16.25 | 28.55 | 33.76 | 37.19 | 17.47 | 29.36 | 34.15 | 36.24 | 19.42 | 18.31 |
| SAHU[25] | 35.22 | 37.76 | 17.86 | 31.25 | 36.93 | 40.68 | 19.16 | 32.12 | 37.34 | 39.64 | 21.28 | 20.08 |
| DK1[26] | 36.16 | 38.66 | 18.79 | 29.38 | 37.95 | 38.37 | 20.42 | 33.19 | 38.32 | 40.57 | 22.29 | 18.69 |
| DK[27] | 35.34 | 37.82 | 18.09 | 28.72 | 36.86 | 37.58 | 19.60 | 32.35 | 37.47 | 39.71 | 21.52 | 18.19 |
| HOROI[28] | 28.08 | 30.16 | 13.87 | 24.76 | 29.36 | 32.41 | 14.87 | 25.36 | 29.58 | 31.45 | 16.51 | 15.54 |
| Denisov[29] | 36.16 | 38.66 | 18.79 | 29.38 | 37.95 | 38.37 | 20.42 | 33.19 | 38.32 | 40.57 | 22.29 | 18.69 |
| VSS1[30] | 34.10 | 36.51 | 17.58 | 29.22 | 35.67 | 38.15 | 18.85 | 31.14 | 36.09 | 38.27 | 20.79 | 18.59 |
| VSS2[31] | 32.77 | 35.04 | 16.86 | 28.19 | 33.90 | 36.61 | 18.41 | 29.97 | 34.63 | 36.68 | 19.91 | 18.19 |
| Royer[32] | 36.08 | 38.63 | 16.89 | 28.70 | 34.85 | 37.56 | 19.93 | 33.01 | 38.26 | 40.56 | 20.14 | 18.15 |
| Royer[33] | 36.09 | 38.61 | 17.39 | 28.85 | 35.89 | 37.52 | 20.13 | 33.07 | 38.26 | 40.54 | 20.74 | 18.57 |
| Royer[34] | 33.68 | 36.05 | 17.04 | 28.55 | 35.61 | 37.36 | 18.73 | 30.85 | 35.72 | 37.85 | 20.40 | 18.07 |
| Royer[35] | 34.79 | 37.24 | 17.40 | 28.71 | 35.66 | 37.56 | 19.27 | 31.84 | 36.89 | 39.10 | 20.72 | 18.17 |
| Brown [36] | 28.06 | 30.14 | 13.74 | 24.70 | 29.34 | 32.41 | 14.74 | 25.31 | 29.56 | 31.43 | 16.39 | 15.42 |
| MB1[37] | 25.81 | 27.72 | 12.29 | 21.87 | 26.59 | 28.94 | 13.52 | 23.20 | 27.10 | 28.81 | 14.64 | 13.28 |
| MB2[38] | 22.50 | 24.19 | 12.78 | 27.10 | 27.75 | 35.62 | 11.63 | 20.15 | 23.58 | 25.09 | 15.23 | 16.84 |
| SemFIS[39] | 33.83 | 36.31 | 16.30 | 28.68 | 34.87 | 37.91 | 18.18 | 30.78 | 35.88 | 38.13 | 19.58 | 17.77 |
| SP[40] | 35.15 | 37.72 | 17.28 | 28.62 | 35.59 | 37.24 | 19.17 | 31.96 | 37.20 | 39.52 | 20.51 | 18.37 |
| AP[41] | 35.71 | 38.25 | 18.83 | 28.74 | 37.77 | 37.60 | 19.65 | 32.47 | 37.65 | 39.95 | 21.93 | 18.20 |
| Akra[42] | 30.22 | 32.34 | 16.83 | 28.51 | 34.05 | 37.02 | 16.81 | 27.67 | 32.03 | 33.95 | 19.99 | 18.43 |
| Dong[43] | 36.08 | 38.63 | 16.89 | 28.70 | 34.85 | 37.56 | 19.93 | 33.01 | 38.26 | 40.56 | 20.14 | 18.15 |

Table 2:Comparison of present work with the experiments [5,13]

| Z | A | Log $T_{expt}$ | *Log* $T_{present}$ |
|---|---|---|---|
| 115 | 287 | -1.46 | -2.16 |
| 115 | 288 | -1.06 | -1.29 |
| 115 | 289 | -0.42 | -0.45 |
| 115 | 290 | -0.62 | -0.54 |
| 116 | 290 | -1.82 | -1.79 |
| 116 | 291 | -1.55 | -2.03 |
| 116 | 292 | -1.74 | -1.43 |
| 116 | 293 | -1.28 | -1.78 |
| 117 | 293 | -1.57 | -1.97 |
| 117 | 294 | -1.11 | -1.97 |
| 118 | 294 | -2.74 | -1.97 |
| 118 | 295 | -3.00 | -4.06 |
| 119 | 290 | -0.39 | -0.57 |
| 120 | 298 | -4.52 | -5.94 |
| 120 | 299 | -4.30 | -5.50 |

Table 2:Comparison of halflives for different decay modes

| Nuclei | α-decay | Fission | Cluster decay<br>$^{9}$Be, $^{10}$B, $^{11}$B , $^{12}$C, $^{14}$N, $^{16}$O, $^{19}$F, $^{20}$Ne, $^{21}$Ne, $^{22}$Ne, $^{23}$Na, $^{24}$Mg, $^{25}$Mg, $^{26}$Mg, $^{27}$Al, $^{28}$Si, $^{29}$Si, $^{30}$Si $^{31}$P $^{32}$S, $^{33}$S, $^{34}$S, $^{35}$Cl, $^{36}$Ar, $^{38}$Ar, $^{40}$Ar, $^{39}$K, $^{41}$K, $^{40}$Ca, $^{42}$Ca, $^{43}$Ca, $^{44}$Ca, $^{46}$Ca | **Decay mode** |
|---|---|---|---|---|
| $^{318}$123 | $2.35\times10^{25}$y | $1.05\times10^{-2}$ s | $10^{25}$ -$10^{70}$y | SF |
| $^{319}$123 | $2.83\times10^{28}$ y | $4.68\times10^{-6}$ s | | SF |
| $^{317}$124 | $2.89\times10^{8}$ y | $1.74\times10^{4}$ y | | SF |
| $^{318}$124 | $3.17\times10^{21}$ y | $2.70\times10^{1}$ y | | SF |
| $^{319}$124 | $1.03\times10^{27}$ y | $2.83\times10^{-2}$ y | | SF |
| $^{320}$124 | $4.80\times10^{30}$ y | $1.91\times10^{-5}$ y | | SF |
| $^{318}$125 | $4.80\times10^{9}$ y | $1.03\times10^{12}$ y | | Alpha |
| $^{319}$125 | $2.83\times10^{22}$ y | $2.46\times10^{9}$ y | | SF |
| $^{320}$125 | $2.35\times10^{27}$ y | $3.81\times10^{6}$ y | | SF |
| $^{321}$125 | $4.38\times10^{29}$ y | $3.99\times10^{3}$ y | | SF |
| $^{319}$126 | $5.77\times10^{11}$ y | $1.00\times10^{20}$ y | | Alpha |
| $^{320}$126 | $3.99\times10^{10}$ y | $3.64\times10^{17}$ y | | Alpha |

Fig. 1: Variation of mass excess of parent (Mp) and daughter ( Md) during alpha decay with mass number (A) for superheavy elements .

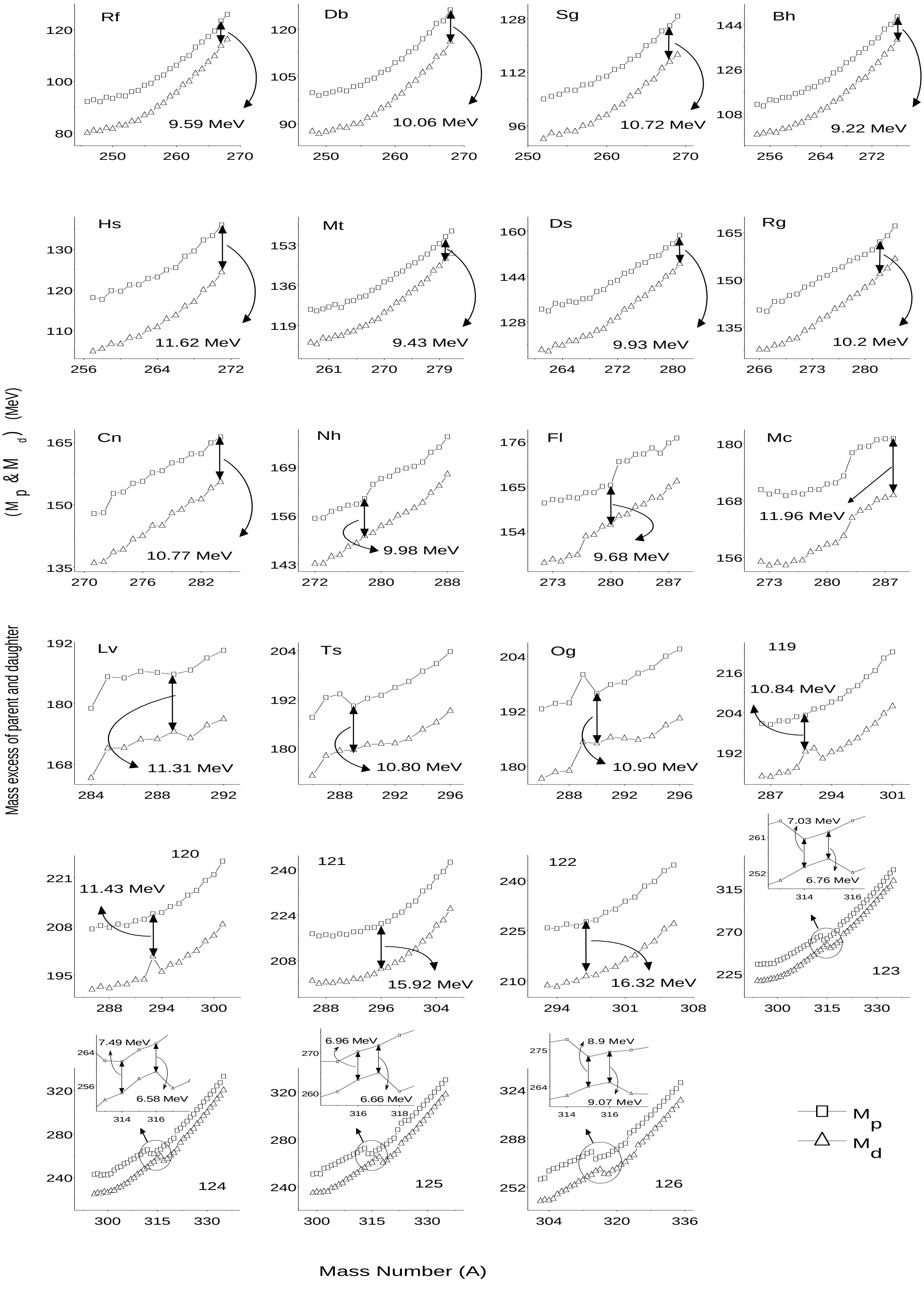

Fig.2: Variation of energy released $Q_\alpha$ (MeV) during alpha decay with mass number (A) for superheavy elements

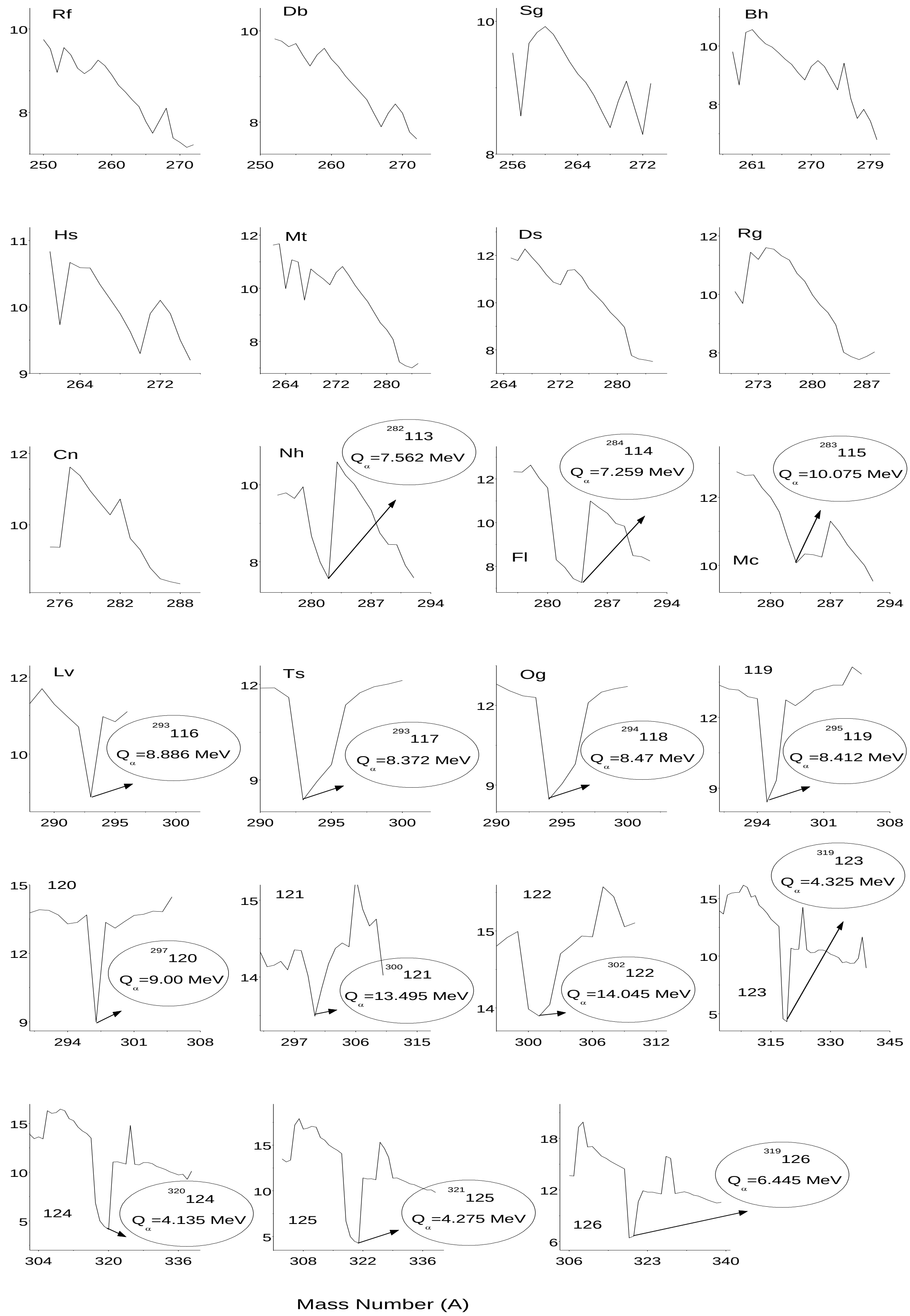

Fig.3: Comparison of spontaneous fission halflives wth that of alpha decay

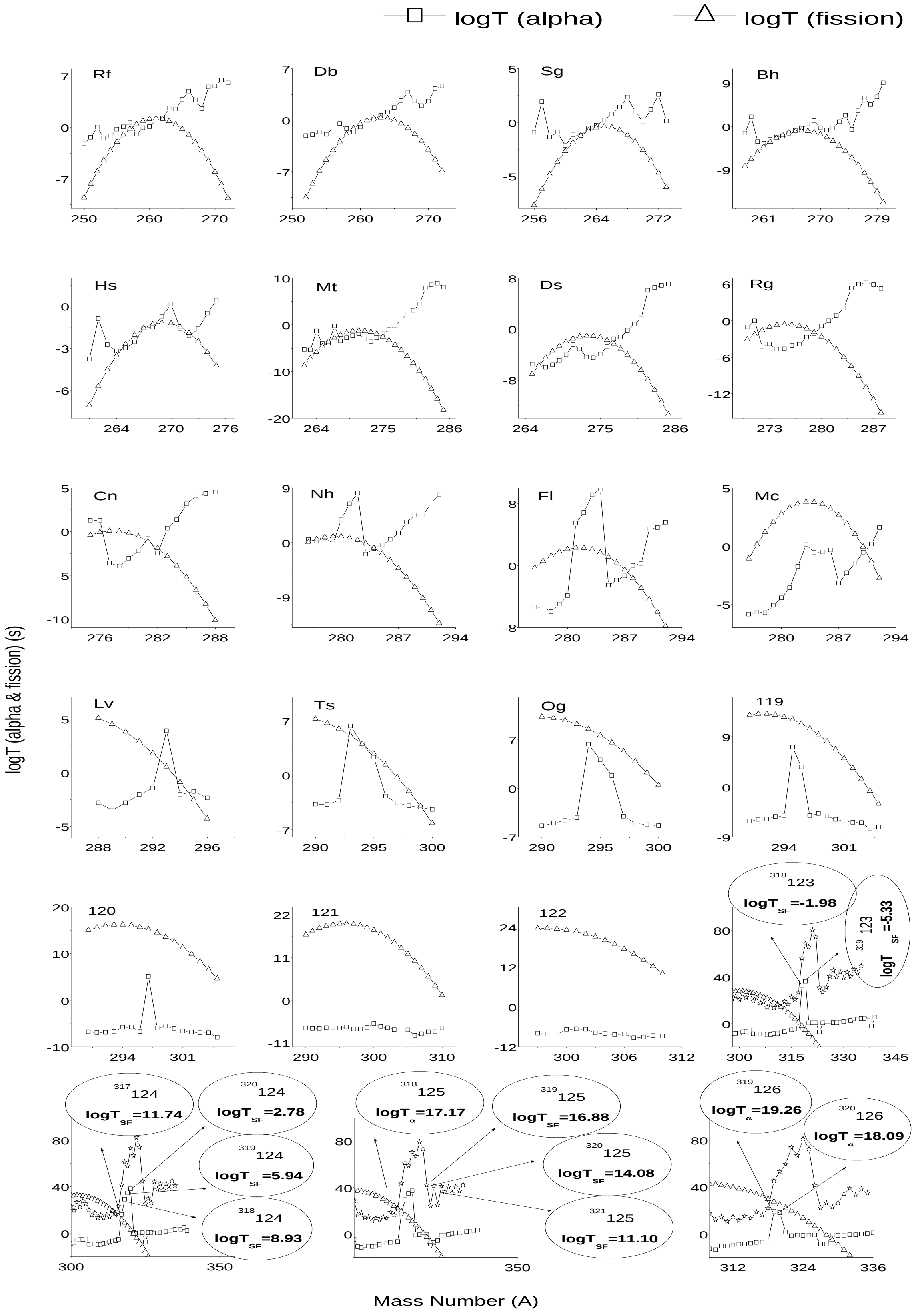